\documentclass[aps,prl,twocolumn,showpacs,graphicx,nofootinbib]{revtex4}

\usepackage{amsmath,amssymb,graphics,graphicx,bbm}

\usepackage{graphics,graphpap}
\usepackage{graphicx}
\usepackage{epsfig}
\usepackage{epstopdf}
\def\roughly#1{\mathrel{\raise.3ex\hbox
{$#1$\kern-.75em\lower1ex\hbox{$\sim$}}}}

\newcommand{\pslash}{D\kern-0.15em\raise0.17ex\llap{/}\kern0.15em\relax}

\begin{document}

\hspace{14cm} \text{MPP--2014--348}

\title{Production of Sterile Neutrino dark matter and the 3.5~keV line}

\author{Alexander Merle$^{a,b}$\footnote{{\tt amerle@mpp.mpg.de}}~~and~~Aurel Schneider$^{c,d}$\footnote{\tt aurel@physik.uzh.ch}}

\affiliation{
${}$\\
\onecolumngrid
$^a${\normalsize \it Max-Planck-Institut f\"ur Physik (Werner-Heisenberg-Institut), F\"ohringer Ring 6, 80805 M\"unchen, Germany}\\
$^b${\normalsize \it Physics and Astronomy, University of Southampton, Southampton, SO17 1BJ, United Kingdom}\\
$^c${\normalsize \it Institute for Computational Science, University of Z\"urich, Winterthurerstrasse 190, CH-8057, Z\"urich Switzerland}\\
$^d${\normalsize \it Department of Physics \& Astronomy, University of Sussex, Brighton, BN1 9QH, United Kingdom}
\twocolumngrid
\onecolumngrid
${}$\\
}

\date{\today}
\begin{abstract}
The recent observation of an X-ray line at an energy of $3.5$~keV mainly from galaxy clusters has initiated a discussion about whether we may have seen a possible dark matter signal. If confirmed, this signal could stem from a decaying sterile neutrino of a mass of $7.1$~keV. Such a particle could make up all the dark matter, but it is not clear how it was produced in the early Universe. In this letter we show that it is possible to discriminate between different production mechanisms with \emph{present-day} astronomical data. The most stringent constraint comes from the Lyman-$\alpha$ forest and seems to disfavor all but one of the main production mechanisms proposed in the literature, which is the production via decay of heavy scalar singlets. Pinning down the production mechanism will help to decide whether the X-ray signal indeed comprises an indirect detection of dark matter.
\end{abstract}

\maketitle

\section{Introduction}
A dream of particle physicists, cosmologists, and astrophysicists is to discover the true nature of dark matter (DM), which makes up more than $80\%$ of the matter in the Universe~\cite{Planck}. The generic candidate is a Weakly Interacting Massive Particle (WIMP), i.e.\ a heavy particle which interacts as weakly as neutrinos. However, the many recent attempts to directly detect such a particle~\cite{DirectDetection} or to produce it at colliders~\cite{LHC-DM}, as well as the hunts for its annihilation products~\cite{IndirectDetection}, have so far not found a clear indication. In this situation, the detection of an X-ray line in several galaxy clusters and in the Andromeda galaxy~\cite{Bulbul2014,Boyarsky2014} has attracted the attention of the community. This line, if stemming from DM decay, could be a smoking gun signal for a very different type of DM particle: an extremely weakly interacting (``sterile'') neutrino with a mass smaller than that of a WIMP by about seven orders of magnitude. WIMPs are produced by \emph{thermal freeze-out}~\cite{Freeze-Out} which means that they decouple from the primordial thermal plasma as soon as the Hubble expansion becomes larger than the interaction rate. Sterile neutrinos with keV-masses cannot be produced in this way because their interactions are too weak. However, even very feebly interacting particles can be gradually produced in the early Universe~\cite{Freeze-In}. For sterile neutrinos this can be achieved by their tiny admixtures $\theta$ to active neutrinos, the so-called Dodelson-Widrow (DW) mechanism~\cite{DW-Mechanism}, but this is known to produce a too hot spectrum~\cite{Boyarsky2009,Ly-alpha_exclusion}, i.e., too fast DM particles. However, active-sterile neutrino transitions could be resonantly enhanced if the background medium carries a net lepton number. This production proposed by Shi and Fuller~\cite{SF-Mechanism} seems in a better shape when confronted with data~\cite{SF-Today}, and it has been recently advocated to be able to produce DM in agreement with the line signal~\cite{Abazajian-Claim}.

What is the status of the 3.5~keV line? Refs. \cite{Bulbul2014, Boyarsky2014} have independently reported evidence in samples by the XMM-Newton and Chandra satellites from nearby clusters and the Andromeda galaxy (stating $>4\sigma$ significance for the stacked signal). These findings were criticised by Refs.~\cite{Riemer-Sorensen:2014yda,Jeltema:2014qfa}, who state that Chandra should see a line signal from the center of the Milky way and that other chemical lines are able to explain the signal. However, these remarks were again criticised in Refs. \cite{Anderson:2014tza,Boyarsky:2014ska,Boyarsky:2014paa,Bulbul:2014ala}, arguing that the centre of the Milky Way is too noisy for a clear signal. Ref.~\cite{Boyarsky:2014ska} questions the range of validity of the background model assumed in~\cite{Jeltema:2014qfa}. Finally, Ref.~\cite{Malyshev:2014xqa} has found no signal in stacked XMM-Newton data from dwarf galaxies, which they claim should provide a clean signal, although the constraint provided is not significantly more stringent than previous ones~\cite{Horiuchi2014}. Obviously the situation is not clear at the moment and more data is required. On the other hand, the technical development of satellites proceeds slower than one would like, so that we cannot expect to see a very bright signal where none had been seen before. Ultimately, we should take the tentative 3.5~keV line as a motivation to scrutinize both the signal and its implications -- the latter we will do here.

With the signal taken seriously, to find out whether DM decay causes it, non-standard production mechanisms must be tested. If the sterile neutrino was charged beyond the SM gauge group~\cite[see][for a review of several such settings]{Merle:2013gea}, freeze-out may be revived if a significant amount of entropy is produced to dilute the abundance~\cite{LR-thermal}, but this is strongly constrained by Big Bang Nucleosynthesis~\cite{keVins}. However, there is another production mechanism which is in a better shape, using a scalar that decays into sterile neutrinos: $S \to \nu_s \nu_s$. This scalar could be the inflaton~\cite{Inflaton-Production} or a general singlet $S$ that is thermally produced in the early Universe by either freeze-out~\cite{WIMP-Production} or freeze-in~\cite{FIMP-Production}. Ultimately the production mechanism has an impact on the DM velocity profile and thus on structure formation.\\

In this letter we present a snapshot of an extensive study to be available soon~\cite{MerleSchneiderTotzauer-Detailed}. We show that, contrary to common believe, sterile neutrino production by scalar decays seems to be in better agreement with data than Shi-Fuller production, in particular when looking at the Lyman-$\alpha$ (Ly-$\alpha$) bound. Knowing which mechanisms fit the data will be of uttermost importance when aiming at identifying whether DM decay could be behind the X-ray signal.

\section{Dark matter production from decays of scalar singlets}
Just as the fermions in the SM obtain their masses by the so-called ``Yukawa'' couplings to the Higgs scalar field $H$, sterile neutrinos $\nu_s$ could couple to a singlet scalar field $\mathcal{S}$ like $\frac{y}{2} \mathcal{S} \overline{\nu_s^c} \nu_s + h.c.$ If $\mathcal{S}$ settles at its vacuum expectation value $v_S = \langle \mathcal{S} \rangle$, this leads to a sterile neutrino mass $m_s = y v_S$ similarly to the ordinary Higgs mechanism. However, the scalar field $\mathcal{S}$ is also allowed by all symmetries to couple to the SM Higgs field via a ``portal'' $H^{\dagger} H \mathcal{S}^2$. This coupling could produce sizeable amounts of $S$ particles (i.e., the physical components of $\mathcal{S}$) which will decay with strength $y$ into two sterile neutrinos. This mechanism can lead to efficient DM production.

Because the $S$ particles only decay efficiently once they are non-relativistic, they do not contribute to the DM momentum distribution. Only their abundance is important for the DM abundance, since every scalar singlet decays in exactly two sterile neutrinos.

The momentum distribution of a decay produced sterile neutrino with adjacent DW production is~\cite{Colombi:1995ze,Kaplinghat:2005sy,Bezrukov:2014qda},
\begin{equation}
f(p)=\frac{\beta_{\rm SD}}{(p/T_{\rm SD})}\exp\left(-p^2/T_{\rm SD}^2\right)+\frac{\beta_{\rm DW}}{\exp(p/T_{\rm DW})+1},
\label{eq:SDdist}
\end{equation}
where $\alpha_{\rm DW}=T_{\rm DW}/T\sim0.716$ and $\alpha_{\rm SD}=T_{\rm SD}/T\sim0.35$ ($T$ being the photon-temperature). A detailed derivation of Eq.~\eqref{eq:SDdist} is given in the Appendix. The normalization factors $\beta_{\rm SD}$ and $\beta_{\rm DW}$ depend on the details of the production mechanism and are fixed by the required DM abundance. This can be determined with the help of the empirical formula from Ref.~\cite{Abazajian2006},
\begin{equation}
\Omega_{\rm DW}\sim7.8\cdot10^{-5}\left[\frac{\sin^2 (2\theta)}{10^{-10}}\right]^{1.23}\left[\frac{m_s}{\rm keV}\right]^{2},
\label{eq:abundance}
\end{equation}
which gives an estimate for the fraction of DM produced non-resonantly via the active-sterile mixing $\theta$.

\begin{figure}
\centering
\includegraphics[trim = 5mm 8mm 32mm 10mm, clip, scale=0.5]{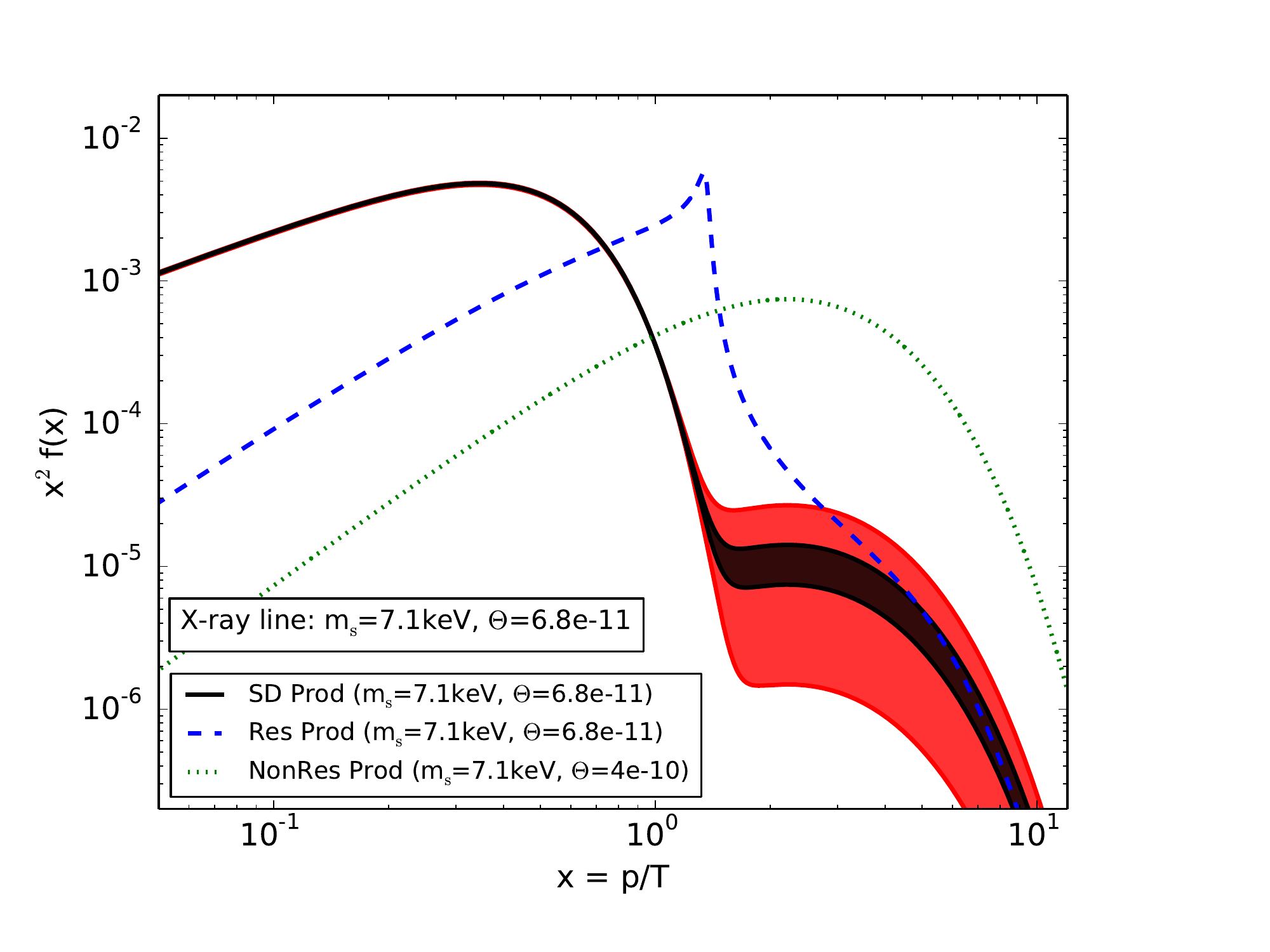}
\caption{\label{Fig:dist}Momentum distribution from scalar decay production (black), resonant production (blue dashed,~\cite{Abazajian-Claim}) and non-resonant production (green dotted). The black and the blue lines use parameters corresponding to the claimed signal~\cite{Bulbul2014}. The green dotted line assumes a larger mixing angle to allow for the right DM abundance. The red envelope around the black line corresponds to the 3$\sigma$ C.L.\ from Ref.~\cite{Bulbul2014}.}
\end{figure}

It becomes clear from Eq.~\eqref{eq:abundance} that the exact form of the momentum distribution depends effectively on two parameters, namely the mass $m_{\rm s}$ of the sterile neutrino and the active-sterile mixing $\Theta \equiv\sin^2 (2\theta)$. Both parameters can be unambiguously determined from the energy spectrum and the flux of the observed X-ray line and are reported to be $m_s=(7.14 \pm 0.07)~{\rm keV}$ and $\Theta=6.8^{+2.2}_{-1.7}\cdot10^{-11}$, respectively~\cite{Bulbul2014,footnote1point5}. The resulting momentum distributions are plotted in Fig.~\ref{Fig:dist}. The black line corresponds to scalar decay production, cf.\ Eq.~\eqref{eq:SDdist}, while the thickness of the line illustrates mixing with different neutrino flavors. The surrounding red band designates the 3$\sigma$ confidence level on the the flux measurement. The distribution exhibits two maxima, one at very cold momenta coming from scalar production and a much smaller one at larger momenta associated with the subdominant active-sterile mixing. The blue dashed line in Fig.~\ref{Fig:dist} shows the momentum distribution resulting from resonant production, calculated in~\cite{Abazajian-Claim}. It assumes a lepton number of $L=4.6\cdot10^{-4}$ and has a characteristic spike at low momenta due to the resonance in the active-sterile mixing. The green dotted line in Fig.~\ref{Fig:dist} illustrates the standard non-resonant production based on the DW mechanism~\cite{DW-Mechanism} as sole source of DM. The X-ray line measurement trivially excludes this mechanism, since non-resonantly produced sterile neutrinos would require a considerably larger mixing angle to make up for all of the DM in the Universe (and they would be too hot). We nevertheless plot the non-resonant case as reference, however, with a mixing angle $\Theta=4\cdot10^{-10}$ to obtain the correct DM abundance.\\

\section{Cosmological Perturbations}

DM particles with a mass in the keV-range are usually categorized as warm DM (WDM) candidates because they generate an important amount of free streaming, which suppresses perturbations at dwarf galaxy scales. The free-streaming length ($\lambda_{\rm fs}$) does however not only depend on the particle mass but also on the average particle momentum, i.e.\ $\lambda_{\rm fs}\sim \langle q\rangle/m_s$. Since the average momentum of scalar-decay produced sterile neutrinos is comparatively small, the effect of free-streaming is expected to be reduced in comparison to production via active-sterile mixing. It is therefore important to properly calculate the free-streaming effect in order to see how strongly scalar decay sterile neutrinos suppress the collapse of dwarf galaxies and whether they act more like warm or cold DM (CDM)~\cite{Petraki:2008ef}.

\begin{figure}
\centering
\includegraphics[trim = 5mm 8mm 32mm 10mm, clip, scale=0.5]{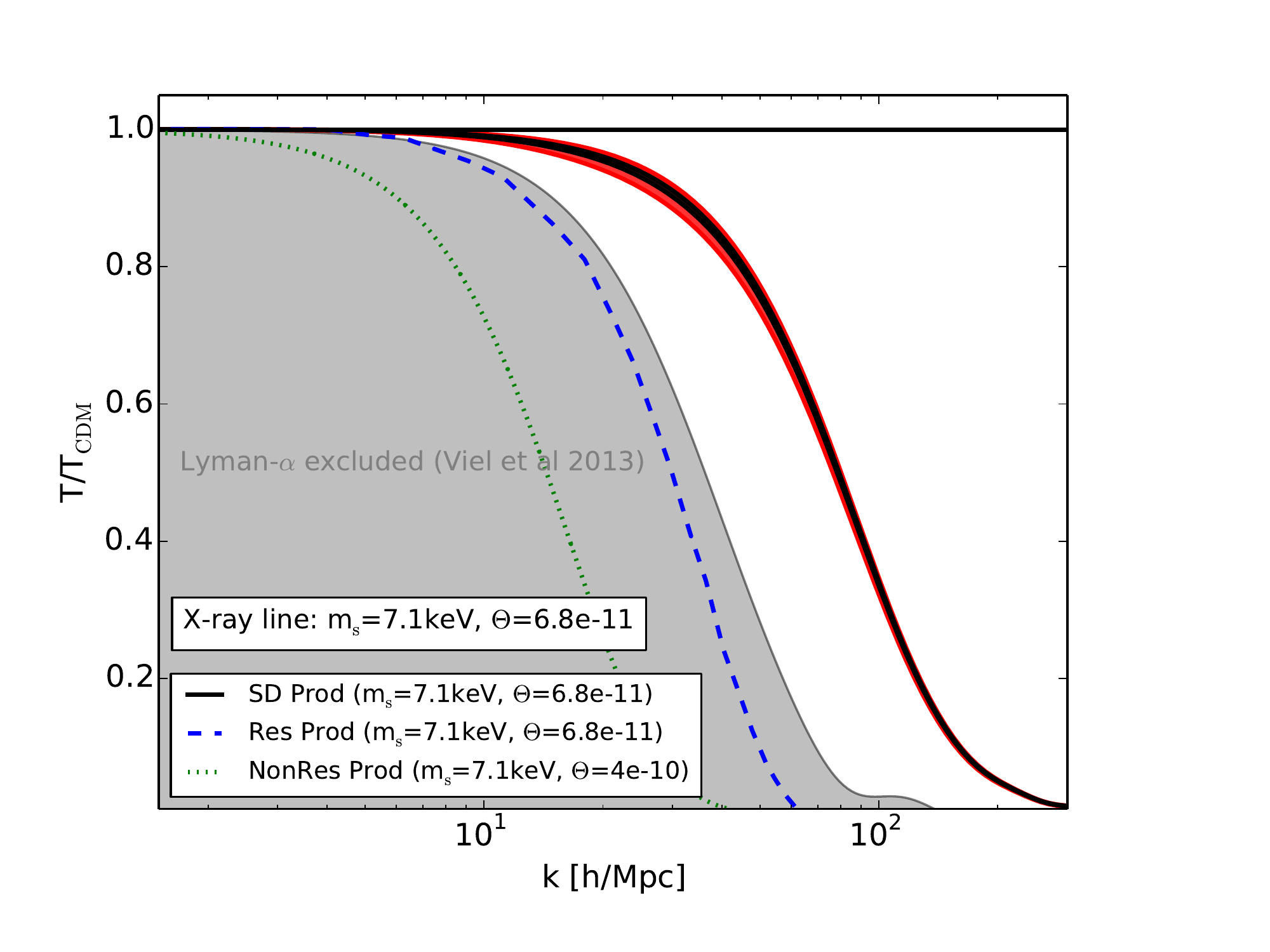}
\caption{\label{Fig:TF}Ratios of the transfer functions (i.e.\ square-root ratios of power spectra, $T/T_{\rm CDM} = \sqrt{P/P_{\rm CDM}}$) from scalar decay production (black, with red-shaded region, corresponding to the 90\%~C.L.), resonant production~\cite{Abazajian-Claim} (blue dashed), and non-resonant production (green dotted). The gray region represents the range of scales disfavored by Ly-$\alpha$~\cite{Viel2013}.}
\end{figure}

We use the numerical Boltzmann solver {\tt CLASS}~\cite{CLASS} to compute matter perturbations for the DM scenarios introduced above. The suppression of small structures with respect to pure CDM is shown in Fig.~\ref{Fig:TF}, where we plot the ratio of the transfer functions (i.e., the square-root of the linear power-spectrum $T/T_{\rm CDM}=\sqrt{P/P_{\rm CDM}}$). The black line with red envelope corresponds to the scalar decay momentum distribution, in agreement with the measured X-ray line and 3$\sigma$ errors~\cite{Bulbul2014}. The blue dashed (green dotted) lines represent (non-)resonant production -- the former is obtained from Ref.~\cite{Abazajian-Claim}. The gray shaded region illustrates the bound on the free-streaming from Ly-$\alpha$ data~\cite{Viel2013,footnote2}.

The transfer functions plotted in Fig.~\ref{Fig:TF} all stem from sterile neutrinos with the exact same mass. The fact that they exhibit very different suppression scales illustrates the strong effect the momentum distribution, and thus the production mechanism, has on particle free-streaming. DM candidates in the keV mass range can either act as cold, warm, or hot DM, depending on their average momentum and their distribution. For this reason it is crucial to know the details of particle production to constrain sterile neutrino DM.

Fig.~\ref{Fig:TF} clearly illustrates the power of the Ly-$\alpha$ method to discriminate different DM scenarios. The non-resonant DW production of sterile neutrinos can be ruled out at high significance~\cite{footnote3}. More surprising is the fact that the resonant production mechanism seems to be in tension with the Ly-$\alpha$ data, too, while the scalar decay production mechanism is perfectly consistent. The tension is at the 2.5$\sigma$ level and hence not strong enough to exclude the resonant scenario as the driving production mechanism. Moreover, there could be sources of error in both data and theory that have to be clarified before drawing any final conclusions. For example, changes in the temperature evolution of the plasma could influence resonant production, or some previously unknown feedback effects could affect the inter-galactic medium (IGM) and therefore the Ly-$\alpha$ analysis ~\cite{IGMresearch}. Another potential source of uncertainty is the X-ray flux measurement. Recent constraints from Ref.~\cite{Horiuchi2014} clearly disfavor mixing angles above $\Theta=5\cdot 10^{-11}$ and taking these more stringent limits into account increases the tension between resonant production and Ly-$\alpha$ data even further (cf.\ Fig.~2 in Ref.~\cite{Abazajian-Claim}).

Despite these potential systematics, it is encouraging that present-day astronomical data can start to discriminate between different production mechanism of sterile neutrino DM. Future weak lensing surveys, such as {\tt EUCLID}, are expected to provide robust constraints and yield an independent check of the Lyman-$\alpha$ results \cite{Markovic2014}, which would form a solid basis to discriminate the known mechanisms.

\section{Halo Formation}
Understanding the formation of DM haloes -- the building blocks of structure formation and main components of every galaxy -- is crucial to distinguish different DM scenarios with astronomical data. Unfortunately, the smallest and most relevant scales are dominated by complex nonlinear physics of \emph{both} gravitational and hydrodynamical origin. The modeling of the hydrodynamical effects is particularly cumbersome because it depends on various feedback mechanisms that are poorly understood and tend to suppress luminous sources, mimicking the expected suppression by WDM.

It has been known for a long time that dwarf galaxy number counts and internal kinematics are in conflict with predictions from $N$-body simulations in the standard $\Lambda$CDM scenario~\cite{SSP}, i.e., the cosmological standard model including Dark Energy and \emph{cold} DM. Suggestions to solve these {\it small scale problems} are numerous and go from invoking a more realistic treatment of baryonic physics~\cite{Hydro} to postulating alternative DM scenarios~\cite{DarkMatter}.

Studying small scale structure formation of sterile neutrino DM in detail would therefore be very desirable. This would however imply running extensive numerical simulations and lies beyond the scope of this work. It is nevertheless possible to gain some insight into nonlinear clustering by applying an extended Press-Schechter (EPS) approach~\cite{EPS}, which approximates structure formation by combining linear growth with idealized ellipsoidal collapse. Standard EPS models are designed for $\Lambda$CDM cosmologies and it turns out that they completely fail in the presence of suppressed power spectra. We therefore use a modified EPS approach constructed to cope with arbitrarily shaped power spectra and tested for warm and mixed DM cosmologies~\cite{sharpk}. In this approach the halo mass function (i.e.\ the number density of haloes per logarithmic mass bin) can be written as
\begin{eqnarray}
&&\frac{dn}{d\ln M} =\frac{\bar\rho}{M}f(\nu)\frac{1}{12\pi^2\sigma^2(R)}\frac{P(1/R)}{R^3}\nonumber\\
&&\ \ {\rm with}\ \ \sigma(R ) =\int \frac{d\mathbf{k}^3}{(2\pi)^3} P(k)\ \Theta_H(1-kR),
\label{eq:massfct}
\end{eqnarray}
where $\Theta_H(x)$ is the Heaviside step-function and $f(\nu)=A\sqrt{2\nu/\pi}(1+\nu^{-p}) {\rm e}^{-\nu/2}$ is the `first-crossing distribution' with $\nu= (1.686/\sigma)^2$, $A=0.322$, and $p=0.3$. The mass and length scales are connected by the relation $M=4\pi\bar\rho(cR)^3/3$, where $c=2.5$. A detailed description of the procedure can be found in Refs.~\cite{sharpk}.

\begin{figure}
\centering
\includegraphics[trim = 5mm 8mm 32mm 10mm, clip, scale=0.5]{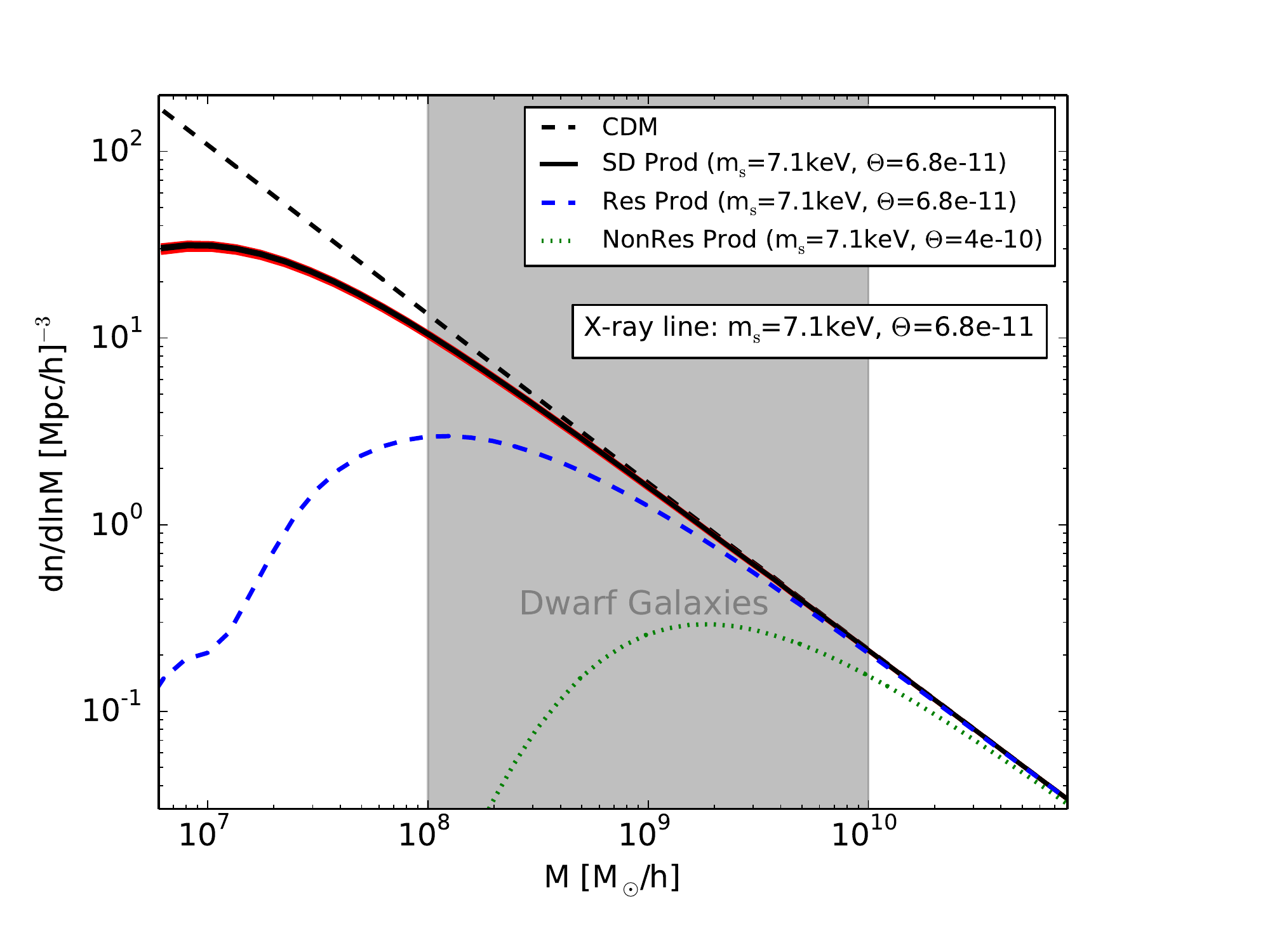}
\caption{\label{Fig:massfct}Halo mass function for WDM from scalar decay production (black, with red-shaded region corresponding to the 3$\sigma$ confidence level), resonant production (blue dashed), and non-resonant production (green dotted). The CDM mass function is given for reference (black dashed line). The gray region illustrates the typical range of scales of dwarf galaxies.}
\end{figure}

In Fig.~\ref{Fig:massfct} we plot the halo mass function based on the transfer functions from Fig.~\ref{Fig:TF}, where the black line (with red band) corresponds to scalar decay production, the blue dashed line to resonant production, and the green dotted line to the standard non-resonant production. For comparison, the CDM halo mass function is given by the dashed black line and the relevant dwarf galaxy scales are highlighted in gray.

This figure shows that the halo mass function of scalar decay produced DM is close to the CDM case at dwarf galaxy scales. The halo abundance starts to be significantly suppressed below a mass of $5\times 10^7 M_\odot$. Since haloes of this mass range are not able to form stars (their gravitational potentials are not deep enough to allow efficient cooling of the gas), the scalar decay scenario is expected to behave very similarly to CDM on astronomically relevant scales. 

The situation is very different for the case of resonant production, where haloes below $10^9 M_\odot$ are strongly suppressed. While this scenario is in tension with Ly-$\alpha$ data, it is expected to alleviate some of the small scale problems of CDM structure formation~\cite{WDMsims}. In the non-resonant scenario, finally, the halo mass function is suppressed in the entire dwarf galaxy range. The suppression is so strong that this scenario is not only ruled out by the Ly-$\alpha$ forest but also because it predicts far fewer Milky-Way satellites than observed~\cite{DwarfGalaxyConstraints}.

\section{Conclusions}
In this letter we have examined the assumption that the recently observed X-ray line in galaxy clusters stems from decays of sterile neutrino DM. In order to check the validity of such a scenario, it is crucial to understand the details of sterile neutrino production in the early Universe. We compared the most prominent production mechanisms and showed that they exhibit considerable differences in the growth of perturbations, which can be distinguished even with \emph{present-day} astronomical observations. The Ly-$\alpha$ signal from high-redshift quasars gives the most stringent constraints and seems to disfavor all but one of the most named sterile neutrino production mechanisms, namely the production via scalar decay, while (non-)resonant oscillations of active into sterile neutrinos seem to be in tension with the Ly-$\alpha$ measurement. Indeed, production via the decay of heavy scalar singlets seems in perfect agreement with the data and could remain as the only valid production mechanism if the $3.5$~keV line observation is solidified.

Additionally, we showed that sterile neutrinos produced from decays at rest have an unusually cold momentum distribution. As a consequence, they are indistinguishable from cold DM at the relevant scales of dwarf galaxies, and do not contribute towards a solution of the highly contested {\it small scale problems} of $\Lambda$CDM.

\section{Acknowledgements}
We thank Kevork Abazajian for useful discussions. AM acknowledges support by a Marie Curie Intra-European Fellowship within the 7th European Community Framework Programme FP7-PEOPLE-2011-IEF, contract PIEF-GA-2011-297557, during his time in Southampton and partial support from the European Union FP7 ITN-INVISIBLES (Marie Curie Actions, PITN-GA-2011-289442). AS is supported by the early researcher fellowship of the Swiss Science Foundation (P2ZHP2\_151605).

\section{Appendix: The phase space distribution function}

In this Appendix, we derive an analytical expression for the momentum (phase space) distribution function for a sterile neutrino DM particle $\nu_s$ produced by the 2-body decay of a scalar $S$ into a pair of $\nu_s$, $S \to \nu_s \nu_s$, which motivates the analytical distribution used in Eq.~\eqref{eq:SDdist}.\\

The Boltzmann equation for the distribution function $f_{\nu_s} (q, t)$ in terms of the comoving momentum $q$ is~\cite{Bernstein}:
\begin{equation}
 \frac{d f_{\nu_s} (q, t)}{d t} = C[q],
 \label{eq:Boltzmann}
\end{equation}
where $C[q]$ is the \emph{collision term}.

The total number of sterile neutrinos, $N_{\nu_s} (t) = R_0^3 n_{\nu_s} (t)$, within a normalized volume $R_0^3$ can be expressed in terms of the number density (which depends on the distribution function),
\begin{equation}
 N_{\nu_s} (t) = R_0^3 \frac{g_{\nu_s}}{(2\pi)^3} \int d^3 q f_{\nu_s} (q, t) \equiv \int\limits_{q=0}^\infty d q\ N^q_{\nu_s} (t),
 \label{eq:Nnus}
\end{equation}
where $g_{\nu_s} = 2$. In the last step we have introduced the \emph{momentum spectral function}
\begin{equation}
N^q_{\nu_s} (t) = \frac{g_{\nu_s}R_0^3}{2\pi^2} q^2 f_{\nu_s} (q, t),
\label{eq:mom_spec}
\end{equation}
which yields the number of particles with momentum $q$ at time $t$. The number of particles produced within given momentum ($[q, q + dq]$) and time ($[t', t' + dt']$) equals the number of DM particles produced per decay, times the number of daughter particles produced within the time interval under consideration (i.e., the negative decay rate of the parent particles times the length of the time interval), times the comoving momentum spectrum $g(q)$ of these particles,
\begin{equation}
 d N^q_{\nu_s} (t') = 2 \times \left( - \frac{d N_S(t')}{dt'} dt' \right) \times g(q),
 \label{eq:derive_mom_spec}
\end{equation}
where $N_S(t') = N_{S,i} e^{-(t' - t_i)/\tau} \simeq N_{S,i} e^{-t'/\tau}$ is the number of parent particles $S$ as a function of time $t'$. $N_{S,i}$ is the initial number of particles at time $t_i \ll t'$, $\tau$ is the lifetime of $S$. In this step we implicitly assume that the parent particles are non-relativistic, since otherwise the lifetime would be modified.

In the rest frame of the parent particle $S$ both sterile neutrinos have the center-of-mass momentum $p_{\rm cm} = M_S/2$, where $M_S$ is the mass of $S$. This momentum is redshifted by the cosmological expansion, so that a particle produced at time $t'$ will at a time $t$ have the physical momentum $p_{\rm cm} a(t')/a(t)$. To compute the corresponding comoving momentum, we should calculate the physical momentum today, $p_{\rm cm} a(t') = q$. The corresponding comoving momentum spectrum for a particle produced at time $t'$ is
\begin{equation}
 g(q) = \delta [q - p_{\rm cm} a(t')].
 \label{eq:g(q)}
\end{equation}
In order to transform this result from the particle rest-frame to the comoving cosmological frame, we need to assume a momentum distribution of the parent particles. An approximate way of doing this is to assume that all parent particles move with the the same momentum $p_{\rm S}$, leading to
\begin{equation}
 g(q) = \delta [q - p_{\rm tot} a(t')],\hspace{0.5cm}p_{\rm tot}=\sqrt{p_{\rm cm}^2+p_S^2},
 \label{eq:g(q)_prime}
\end{equation}
i.e a similar correction than the one used in Ref.~\citep{Bezrukov:2014qda}.

Using $- d N_S(t')/dt' \simeq N_{S,i} e^{-t'/\tau} / \tau$, we obtain the relation
\begin{equation}
 \frac{d N^q_{\nu_s} (t')}{dt'} = \frac{2 N_{S,i}}{\tau} e^{-t'/\tau} \delta [q - p_{\rm tot}a(t')].
 \label{eq:derive_mom_spec_final}
\end{equation}
which can be integrated over the interval $\left[t_i \simeq 0, t\right]$ to obtain the momentum spectral function
\begin{equation}
 N^q_{\nu_s} (t) = \frac{2 N_{S,i}}{\tau} \int\limits_{t'=0}^\infty dt'\ e^{-t'/\tau} \delta [q - p_{\rm tot} a(t')].
 \label{eq:dmom_spec_t}
\end{equation}
Since the DM production happens during the radiation dominated regime, one can make use of $a(t) \propto t^{1/2}$ to derive
\begin{eqnarray}
 N^q_{\nu_s} (t) &=& \frac{4 N_{S,i} q t}{\tau p_{\rm tot}^2 a^2(t)}\times\label{eq:dmom_spec_t_expl}\\
 &&\exp\left[ - \frac{q^2}{p_{\rm tot}^2 a^2(t)} \frac{t}{\tau} \right] \Theta_H [p_{\rm tot} a(t) - q].\nonumber
\end{eqnarray}
For the limiting case of $p_{\rm tot}=p_{\rm cm}$ (i.e.\ vanishing momenta of the parent particles), this result exactly coincides with what has been obtained in Refs.~\cite{Kaplinghat:2005sy,Bezrukov:2014qda}.

The distribution function can now be obtained by combining the Eqs.~\eqref{eq:dmom_spec_t_expl} and \eqref{eq:mom_spec}. Let $N_{S,i} / R_0^3 \equiv n_{S,i}$ be the initial number density of parent particles $S$ inside the comoving volume~\cite{footnote4}. This allows to write the physical distribution function as
\begin{equation}
 f_{\nu_s} (p, t) = \frac{\beta}{(p/T_{\rm SD})} e^{-p^2 / T_{\rm SD}^2} \ \ \Theta_H \left[T_{\rm SD}\frac{a(t)}{a(\tau)} - p\right],
 \label{eq:distribution_comoving}
\end{equation}
where we have defined normalization and temperature parameters:
\begin{equation}
\beta \equiv \frac{2 \pi^2}{g_{\nu_s}}\frac{n_{s,i}}{[a(t)T_{\rm SD}]^{3}},\hspace{0.5cm}T_{\rm SD} \equiv p_{\rm tot} \frac{a(\tau)}{a(t)}.
\label{eq:parameters}
\end{equation}
In order to fully recover the first part of Eq.~\eqref{eq:SDdist} in the main text, we still have to connect $T_{\rm SD}$ to the photon temperature $T$. This can be achieved by determining $p_{\rm tot}$ and $a(\tau)$ in Eq.~\eqref{eq:parameters}. Since every parent particle decays into 2 sterile neutrinos (i.e.\ $p_{\rm cm}=M_{\rm S}/2$) as soon as it becomes non-relativistic (i.e.\ $p_S=M_S$), the total momentum at decay is 
\begin{equation}
p_{\rm tot}=\sqrt{p_{\rm cm}^2+p_S^2}\simeq\sqrt{5/4}\ M_S.
\label{eq:ptot}
\end{equation}
Furthermore, entropy conservation ($g_{e\rm ff}(Ta)^3=\rm const.$) yields
\begin{equation}
\frac{a(\tau)}{a(t)}=\left(\frac{g_{\rm eff,0}}{g_{\rm eff,\tau}}\right)^{1/3}\frac{T(t)}{T(\tau)}\simeq\left(\frac{g_{\rm eff,0}}{g_{\rm eff,\tau}}\right)^{1/3}\frac{T(t)}{M_S}.
\label{eq:atau}
\end{equation}
Combining Eqs.~\eqref{eq:parameters}, \eqref{eq:ptot}, and \eqref{eq:atau} with the values $g_{\rm eff,\tau}=109.5$ and $g_{\rm eff,0}=3.36$ finally leads to the ratio
\begin{equation}
\alpha_{\rm SD}=\frac{T_{\rm SM}}{T}\simeq0.35
\end{equation}
used in the main text.


\begin{thebibliography}{76}
\bibitem{footnote1point5}
For the latter value we have taken the mean between the 1$\sigma$ ranges of both MOS and PN instruments.
\bibitem{footnote2}
The Ly-$\alpha$ constraint was derived using a renormalized Fermi-Dirac distribution resulting from non-resonant production~\cite{DW-Mechanism}. Technically, only models with the same distribution can be compared to this constraint, while other models would require an independent analysis of the nonlinear clustering. In practice, however, all models with transfer functions predominantly lying in the gray area are expected to be excluded by Ly-$\alpha$, independent of the exact shape of their spectra.
\bibitem{footnote3}
This should not come as a surprise, since the scenario is equally at odds with Dwarf Galaxy observations from the Milky Way~\cite{DwarfGalaxyConstraints}.
\bibitem{footnote4}
The only change in the number density $n_S(t)$ of $S$ within the comoving volume comes from decays, since the expansion is mapped out. Because each $S$ produces exactly two $\nu_s$, the final comoving number density of sterile neutrino is twice that of the initial parent particles, $n_{\nu_s} (t \to \infty) = 2 n_{S,i}$, as soon as the production of $S$ is finished.
\end{thebibliography}
\end{document}